\documentclass{article}
\usepackage{graphicx} 
\usepackage{subfigure}
\usepackage{amsfonts}       
\usepackage{breqn}
\usepackage{url}    
\usepackage{xcolor}

\newcommand{\beq}{\begin{eqnarray}}
\newcommand{\eeq}{\end{eqnarray}}
\newcommand{\bpmatrix}{\begin{pmatrix}}
\newcommand{\epmatrix}{\end{pmatrix}}
\newcommand{\ba}{\begin{array}}
\newcommand{\ea}{\end{array}}

\def\bea{\begin{eqnarray}}   
\def\eea{\end{eqnarray}}

\title{Invisible dark matter decays of a non-Standard Model like CP-even scalar boson}

\author{
 Maien Binjonaid \\
  Department of Physics and Astronomy\\
  King Saud University\\
  Riyadh, Saudi Arabia \\
  \texttt{maien@ksu.edu.sa} \\
}

\begin{document}

\maketitle

\begin{abstract}
We investigate two extensions of the standard model that include particle dark matter candidates: the Next-to-Two Higgs Doublet Model and the Next-to-minimal Supersymmetric Standard Model. These models feature a non-Standard Model like CP-even scalar with a sub-TeV mass, denoted by $H_2$, among other particles. At a 13 TeV proton-proton collider, the primary production channel for such scalars is via the fusion of a pair of gluons. Subsequently, these scalars can decay invisibly into a pair of dark matter candidates, which can be dominant. In the supersymmetric model, it is possible for the Lightest Supersymmetric Particle (LSP) and Next-to Lightest Supersymmetric Particle (NLSP) to be mass degenerate, leading to quasi-invisible $H_2$ decays to LSP+NLSP and NLSP+NLSP. We present the predictions of both models for this challenging scenario while ensuring compatibility with recent experimental constraints.

\end{abstract}
\section{Introduction} \label{intro}

The discovery of the Standard Model (SM) like Higgs boson at the Large Hadron Collider (LHC) \cite{Aad:2012tfa, Chatrchyan:2012xdj} marked the completion of the SM of particle physics, and opened the door for what could lay beyond it, which is usually called Beyond the SM (BSM) phenomenology. The SM suffers from multiple shortcomings, such as lacking a Dark Matter (DM) candidate, the issue of neutrino mass, the lightness of the Higgs mass (the hierarchy problem), and other issues. The limitations of the SM motivate the construction of several extensions. For instance, a well-known extension is the Minimal-Supersymmetric Standard Model (MSSM), which solves the hierarchy problem and provides a DM candidate. However, given the absence of supersymmetry (SUSY) at collider searches, it is possible to consider a case where most SUSY particles reside at high scales beyond the reach of current experiments. What remains could include the extended Higgs sector of such models. A general non-SUSY version would be the well-known Two-Higgs-Doublet Model (2HDM), which allows for broader Higgs sector structures and Yukawa couplings.  

However, both the 2HDM and the MSSM have their own issues. The 2HDM, in its simplest form, where both Higgs doublets acquire Vacuum Expectation Values (VEVs) does not contain a DM candidate. As for the MSSM, it contains a SUSY Higgs/Higgsino mass term $\mu$ that respects all symmetry conditions, but it is unclear at which scale it is generated, which is known in the literature as the "$\mu$-problem". Solving the latter issue leads us naturally to the Next-to-Minimal Supersymmetric Model (NMSSM), which features an additional SM singlet. A non-SUSY version of such a model is the Next-to-2-Higgs Higgs Doublet Model (N2HDM). Both models are subject to intensive research as they provide DM candidates and novel phenomenology relevant to LHC and DM searches.

The Higgs sectors of the N2HDM and NMSSM comprise an SM-like Higgs ($h$), CP-even and odd neutral scalars ($H_1, H_2, H_3/A_1, A_2$), and charged Higgses ($H^{\pm}$). comprehensive reviews of both models are given in Ref.~\cite{M_hlleitner_2017} and Ref.~\cite{Ellwanger:2009dp}. In these models, the DM particle can be an SM singlet. Furthermore, it is possible for the non-SM Higgs bosons to decay invisibly to DM, making them difficult to detect if such decays are dominant. Hence, understanding the size and dominance of these decays in different parameter regions could be relevant for LHC searches.

A number of papers considered different types of the N2HDM with either real or complex singlets. For instance, the effects of the mixing between the doublets and a singlet with $\langle s \rangle \neq 0$ are studied in Ref.~\cite{Chen:2013jvg}, while in Ref.~\cite{Drozd:2014yla} the model, with a DM candidate, was confronted with relevant experimental constraints at the time. Comparing the cases with a real and complex singlet added to the 2HDM is provided in Ref.~\cite{Dutta:2022xbd}, and it was noted that the case with a real singlet provides larger values of the dark matter relic abundance ($\Omega h^2$) throughout the parameter space.

On the other hand, the NMSSM is a much richer model given the number of new SUSY particles and its connection to the Grand Unification (GUT) scale. The phenomenology of the Higgs sector in different variants of the NMSSM is considered in e.g. Refs.~\cite{Miller:2003ay, Wang:2020tap, Baum:2019uzg,King:2014xwa, Ellwanger:2012ke, Telba:2021gva,Telba:2020pji}. Different types of DM candidates, including the case with singlet DM, are considered in e.g. Refs.~\cite{Wang:2020dtb, Domingo:2022pde, Cao:2022ovk, Ellwanger:2014hia, PhysRevD.108.035020}. The case where the DM is singlino-like requires a specific setup in which the parameters $\lambda$ and $\kappa$ and the ratio $\kappa / \lambda$ are quite small. In this work, we do not specifically target this scenario, as we are allowing $\lambda$ and $\kappa$ to be as large as possible. It is known in the NMSSM that a large value of $\lambda$ enhances the tree-level mass of the SM-Higgs boson, although it still needs to be below 0.7 to abide by GUT scale perturbativity.

From a phenomenological point of view, it is crucial to analyze the differences between models. Indeed comparing some aspects of the N2HDM with the NMSSM was performed in different studies. For example, Refs.~\cite{Muhlleitner:2017dkd, Azevedo:2018llq} show that it is possible to use the couplings sums of the Higgs boson to $VV$ and $f\bar{f}$ to distinguish these models at the LHC, and that future $e^+ e^-$ colliders can be used to do so by setting limits on the possible singlet or pseudoscalar admixtures to the SM Higgs. The possibility of explaining certain anomalies in the LHC data using the N2HDM and the NMSSM is considered in Ref.~\cite{Biekotter:2021qbc}.

As mentioned earlier, the additional non-SM CP-even scalar field $H_2$ could have dominant decays into DM, hence making the task of observing it very challenging. Within the context of the MSSM, Ref.~\cite{anan15} investigates invisible decays of $H_2$. The analysis includes the case where $H_2$ decays into the Lightest Supersymmetric Particle (LSP), which is the DM particle, and Next-to LSP (NLSP) with degenerate mass. Such decays are called "quasi-invisible". For the specific parameter space, it was found that $Br(H_2 \rightarrow \tilde{\chi}_1 \tilde{\chi}_1)$ is $16\%$ at most. As for the NMSSM, and as far as we know, no dedicated analysis was performed for such cases, nor a comparison to the N2HDM was made.

We fill this gap, limiting our analysis to the second lightest CP-even non-SM-like Higgs ($H_2$). Firstly, we will ask what are the maximum branching ratios of $H_2$ into DM given the latest experimental constraints. Second, we study the production of $H_2$ via gluon fusion at a 13 TeV pp collider and calculate the cross-section times the branching ratio. The paper is organized as follows. In Sec.~\ref{sec:core1}, we present a general overview of the two considered models, their input parameters, Higgs sectors, and relevant quantities. Next, Sec.~\ref{sec:scans} details the tools utilized in this paper and the constraints applied in analyzing the parameter space of each model. In Sec.~\ref{sc:res}, the results for both models are presented. And finally, the discussion and conclusions are given in Sec.~\ref{discon}.

\section{The models}
\label{sec:core1}

\subsection{Overview of the N2HDM}
\label{core:n2hdm}

The N2HDM comprises two-Higgs-doublets ($\Phi_1$ and $\Phi_2$) and a real singlet ($\Phi_S$), and its potential reads (following the conventions in Ref.~\cite{M_hlleitner_2017}),
\beq
V &=& m_{11}^2 |\Phi_1|^2 + m_{22}^2 |\Phi_2|^2 - m_{12}^2 (\Phi_1^\dagger
\Phi_2 + h.c.) + \frac{\lambda_1}{2} (\Phi_1^\dagger \Phi_1)^2 +
\frac{\lambda_2}{2} (\Phi_2^\dagger \Phi_2)^2 \nonumber \\
&& + \lambda_3
(\Phi_1^\dagger \Phi_1) (\Phi_2^\dagger \Phi_2) + \lambda_4
(\Phi_1^\dagger \Phi_2) (\Phi_2^\dagger \Phi_1) + \frac{\lambda_5}{2}
[(\Phi_1^\dagger \Phi_2)^2 + h.c.] \nonumber \\
&& + \frac{1}{2} m_S^2 \Phi_S^2 + \frac{\lambda_6}{8} \Phi_S^4 +
\frac{\lambda_7}{2} (\Phi_1^\dagger \Phi_1) \Phi_S^2 +
\frac{\lambda_8}{2} (\Phi_2^\dagger \Phi_2) \Phi_S^2 \;,
\label{eq:n2hdmpot}
\eeq
where the mass parameters $m_{ij}$ ($i,j=1,2$ and $i \leq j$) correspond to the doublets, while $m_S$ is the mass parameter of the singlet, and the $\lambda$'s are quartic couplings between the scalar fields. Moreover, the structure of the potential in Eq.\ref{eq:n2hdmpot} respects two discrete symmetries. The first is a $\mathbb{Z}_2$ symmetry similar to that in the 2HDM, under which $\Phi_1$ and $\Phi_S$ are even, while $\Phi_2$ is odd. The second is a new $\mathbb{Z}^{\prime}_2$ symmetry, under which $\Phi_1$ and $\Phi_2$ are even, while $\Phi_S$ is odd. This $\mathbb{Z}^{\prime}_2$ symmetry is responsible for the existence of a dark matter candidate in the theory if $\Phi_S$ does not acquire a VEV.  

For the case where the $\mathbb{Z}^{\prime}_2$ symmetry is intact, only the Higgs doublets acquire VEVs. This case is named "the Dark Singlet Phase (DPS)" by the authors of Ref.~\cite{M_hlleitner_2017}, and is the one we are considering (we will denote the model in the subsequent sections by DSP-N2HDM).
Moreover, the vacuum structure is,
\begin{align}
\left<\Phi_1\right>=\dfrac{1}{\sqrt{2}}\begin{pmatrix}
0 \\ v_1
\end{pmatrix},\qquad
\left<\Phi_2\right>=\dfrac{1}{\sqrt{2}}\begin{pmatrix}
0 \\ v_2
\end{pmatrix},\qquad
\left<\Phi_S\right>=0\,,
\end{align}
where $v_1\equiv v\cos\beta$, and $v_2\equiv v\sin\beta$, while $\tan{\beta} \equiv \frac{v_2}{v_1}$, and $v = \sqrt{v_1^2 + v_2^2}$. 
The gauge eigenstates are rotated into mass eigenstates via a rotation matrix $\mathcal{R}$ that depends on the rotation angle $\alpha$. The rotation is done such that $m_{H_1} \le m_{H_2}$. On the other hand, the dark singlet scalar is $H_3 \equiv H_D$, as it does not mix with the other Higgs scalars.

The couplings of $H_1$ and $H_2$ to SM particles are the same as in Type-I 2HDM. For $H_2$, which is our focus in this paper, $c(H_2 \bar{f}f) = \sin\alpha/\sin\beta$, while $c(H_2 VV) = \cos\left(\alpha-\beta\right)$.
In addition to these couplings, we have the triple-Higgs couplings that are responsible for the possible decays of the CP-even Higgs scalars into DM. Particularly,
\begin{dmath}
    g(H_2 H_D H_D) = \lambda_7 \nu \cos{\beta} \cos{\alpha} + \lambda_8 \nu \sin{\beta} \sin{\alpha} 
\end{dmath}

Finally, the DSP-N2HDM is represented by 11 input parameters,
\begin{alignat}{2}
v\,,\enspace \tan\beta\,,\enspace
m_{H_{1}}\,,\enspace m_{H_{2}}\,,\enspace m_{H_{D}}\,,\enspace
m_{A}\,,\enspace m_{H^{\pm}}\,, \alpha\,,\enspace \lambda_6\,,\enspace \lambda_7\,,
\enspace \lambda_8\,,\notag
\end{alignat} 
where $m_{H_1}$ is taken to be the mass of the SM Higgs boson, while $m_{H_D}$ is the mass of the DM candidate, $m_{A}$ is the mass of the CP-odd Higgs boson, $m_{H^{\pm}}$ is the mass of the charged Higgs bosons. The other parameters were defined earlier. 
\subsection{Overview of the NMSSM}
\label{core:nmssm}
The NMSSM is a well-known SUSY model that solves the $\mu-$problem in the MSSM due to adding a singlet superfield $S$. This singlet  couples to the Higgs doublets, extending the well-known MSSM superpotential. Following the convention in Ref.~\cite{Ellwanger:2009dp}, the superpotential is, 
\begin{eqnarray}
W_{NMSSM} &=&h_{u}\widehat{Q}.\widehat{H}_{u}\widehat{U}_{R}^{c}+h_{d}%
\widehat{H}_{d}.\widehat{Q}\widehat{D}_{R}^{c}+h_{e}\widehat{H}_{d}.\widehat{%
L}\widehat{E}_{R}^{c}  \nonumber \\
&&+\lambda \widehat{S}\widehat{H}_{u}.\widehat{H}_{d}+\frac{1}{3}\kappa 
\widehat{S}^{3}.
\end{eqnarray}
where $\widehat{Q}$ and $\widehat{L}$ are
left-handed doublet quark and lepton superfields, whereas $\widehat{U}$, $%
\widehat{D}$ and $\widehat{E}$ are right-handed singlet up-type
quark, down-type quark, and lepton superfields. Unlike the N2HDM, the NMSSM singlet superfield obtains a VEV $\left\langle S\right\rangle =s$.
To avoid a massless Axion at the weak scale, the last term is introduced \cite{PhysRevLett.38.1440, PhysRevD.16.1791}.

Furthermore, the soft SUSY breaking lagrangian of the NMSSM is, 
\begin{eqnarray}
-\mathcal{L}_{soft} &=&m_{H_{u}}^{2}\left\vert H_{u}\right\vert
^{2}+m_{H_{d}}^{2}\left\vert H_{d}\right\vert ^{2}+m_{S}^{2}\left\vert
S\right\vert ^{2}+m_{Q}^{2}\left\vert Q^{2}\right\vert   \nonumber \\
&&+m_{U}^{2}\left\vert U_{R}^{2}\right\vert +m_{D}^{2}\left\vert
D_{R}^{2}\right\vert +m_{L}^{2}\left\vert L^{2}\right\vert
+m_{E}^{2}\left\vert E_{R}^{2}\right\vert   \nonumber \\
&&+\frac{1}{2}\left[ 
\begin{array}{c}
M_{1}\lambda _{1}\lambda _{1}+M_{2}\sum_{i=1}^{3}\lambda _{2}^{i}\lambda
_{i2} \\ 
+M_{3}\sum_{a=1}^{8}\lambda _{3}^{a}\lambda _{a3}%
\end{array}%
\right]   \nonumber \\
&&+h_{u}A_{u}Q\cdot H_{u}U_{R}^{2}-h_{d}A_{d}Q\cdot
H_{d}D_{R}^{2}-h_{e}A_{e}L\cdot H_{d}E_{R}^{2}  \nonumber \\
&&+\lambda A_{\lambda }H_{u}\cdot H_{d}S+\frac{1}{3}\kappa A_{\kappa
}S^{3}+h.c.
\end{eqnarray}
which comprises terms for scalar mass parameters, gaugino mass parameters, trilinear couplings, and dimensionless couplings.

The non-SM CP-even Higgs couplings to two neutralinos (with a focus on the NLSP and the LSP) is,
\begin{align}
H_2 \chi^0_i \chi^0_j &: \frac{\lambda}{\sqrt{2}} (S_{21} \Pi_{ij}^{45} +
S_{22} \Pi_{ij}^{35} + S_{23} \Pi_{ij}^{34}) - \sqrt{2} \kappa S_{23}
N_{i5} N_{j5} \nonumber \\ 
&\quad + \frac{g_1}{2} (S_{21} \Pi_{ij}^{13} - S_{22}
\Pi_{ij}^{14}) - \frac{g_2}{2} (S_{21} \Pi_{ij}^{23} - S_{22}
\Pi_{ij}^{24}).
\end{align}
where $S_{ij}$ represent elements of the Higgs mixing matrix, and $\Pi_{ij}^{ab} \equiv N_{ia}N_{jb}+N_{ib}N_{ja}$ are defined in terms of the mixing matrices of the neutralinos.

Finally, the input parameters of the NMSSM, which are specified at the GUT scale, are $$
m_0, \ m_{1/2}, \ A_{0}, \ A_{\lambda }, \ 
A_{\kappa }, \ \lambda ,\ \kappa , \ \tan{\beta},  \  \mu _{\text{eff}}. $$
comprising scalar and gaugino mass parameters, trilinear couplings, the singlet-Higgs couling$\lambda$, the cubic singlet self-coupling $\kappa$, and $\mu_{\text{eff}} \equiv \lambda s$. Both $m_{H_u}$ and $m_{H_d}$ can have the same value as $m_0$ at the GUT scale, but the case where they differ from $m_0$ is called the Non-Universal Higgs NMSSM (NUH-NMSSM), which we consider here. However, due to the nature of the tool we use (described in Sec.~\ref{sec:scans}) these two parameters are computed.

\section{The parameter spaces} \label{sec:scans}

\subsection{The DSP-N2HDM}
To generate the mass-spectrum generator of the N2HDM we use $\mathtt{N2HDECAY}$ \cite{Engeln:2018mbg}, and is embedded in the state-of-the-art scanning tool $\mathtt{ScannerS}$ \cite{scanners20}. Specifically, we consider the dark-singlet phase of the N2HDM and scan the parameter space over the following ranges:
\begin{eqnarray*}
& m_{H_2}=[10-1500] GeV,
& m_{H_D}=[1-1500] GeV,  \\
& m_{A}=[1-1500] GeV, 
& m_{H^{\pm}}= [150-1500] GeV,  \\
& \alpha =[-1.5-1.6], 
& \tan{\beta} = [0.8-25], \\
& m_{12}^2=[1\times 10^{-3}- 5 \times 10^5 ] GeV^2, \\
& \lambda_6 = [0-20], 
& \lambda_{7,8} = [-30-30].
\end{eqnarray*}
The package applies stringent theoretical and experimental constraints
on the model. The former includes perturbative unitarity, boundedness, vacuum stability, constraints from electroweak precision, and flavor conservation. The latter include Higgs searches and measurements via interfacing with $\mathtt{HiggsBounds}$ $\mathtt{(v.5.9)}$ \cite{Bechtle:2020pkv} and $\mathtt{HiggsSignals}$ $\mathtt{(v.2.6)}$ \cite{Bechtle:2020uwn}, and Dark matter constraints using $\mathtt{MicrOMEGAs}$ $\mathtt{(v.5)}$ \cite{Belanger:2020gnr}. It is worth mentioning that HiggsBounds provides the production cross-section of $H_2$ via gluon fusion at 13 TeV via tabulated results from $\mathtt{SusHi 1.6.1}$ \cite{Harlander:2016hcx}.

\subsection{The NUH-NMSSM}
The parameter space of the NMSSM was scanned using $\mathtt{NMSSMTools}$ $\mathtt{(v.5.6)}$ \cite{NMSSMTools,Ellwanger:2004xm,Ellwanger:2005dv,Das:2011dg}. A combination of random and Markov Chain Monte Carlo (MCMC) sampling methods was deployed. 
The scanned ranges of the input parameters are:
\begin{eqnarray*}
& m_0=[1-4000] GeV, 
& m_{1/2}=[1-4000] GeV,  \\
& A_0,A_{\lambda},A_{\kappa}=[-3000-3000] GeV, 
& \tan{\beta}= [1-30]  \\
& \lambda,\kappa =[0.01-0.7], 
& \mu_{eff}=[100-1500] GeV.
\end{eqnarray*}
As mentioned in Sec.~\ref{core:nmssm}, the parameters $m_{H_u}$ and $m_{H_d}$ are computed, and hence what we are considering here is the NUH-NMSSM.
The constraints implemented in $\mathtt{NMSSMTools}$ are similar to those in $\mathtt{ScannerS}$ if not more comprehensive (see the tool's History Section for all constraints included and the corresponding references). 
These include non-tachyonic masses, successful Electroweak Symmetry Breaking, and the existence of a global minimum, which represents the theoretical constraints. In contrast, the phenomenological ones include: Flavor physics, LHC constraints
on sparticles, satisfying the upper limit on dark matter relic density using $\mathtt{MicrOMEGAs}$ $\mathtt{(v.5)}$. As for the SM-Higgs couplings, we consider recent results from the LHC, and select points to be within $3 \sigma$ away from the central values of the combination of ATLAS  and CMS results \cite{CMS:2022dwd, ATLAS:2022vkf}, which makes our results subject to the latest LHC limits. Finally, the production cross-section of $H_2$ via gluon fusion is calculated through the method described in Ref.~\cite{Ellwanger:2022jtd}, where the data for the BSM production cross-section at 13 TeV is obtained from Ref.~\cite{CERNYellow} and subsequently multiplied by the relevant reduced coupling squared. Thus, the calculation accounts for a significant portion of the radiative Quantum Chromodynamics corrections, leaving theoretical uncertainties of the order of $O(10\%)$.

\section{Results} \label{sc:res}

\subsection{The DSP-N2HDM}
\label{res:n2hdm}
\begin{figure}[!ht]
    \centering
    \subfigure[]{\includegraphics[width=0.45\textwidth]{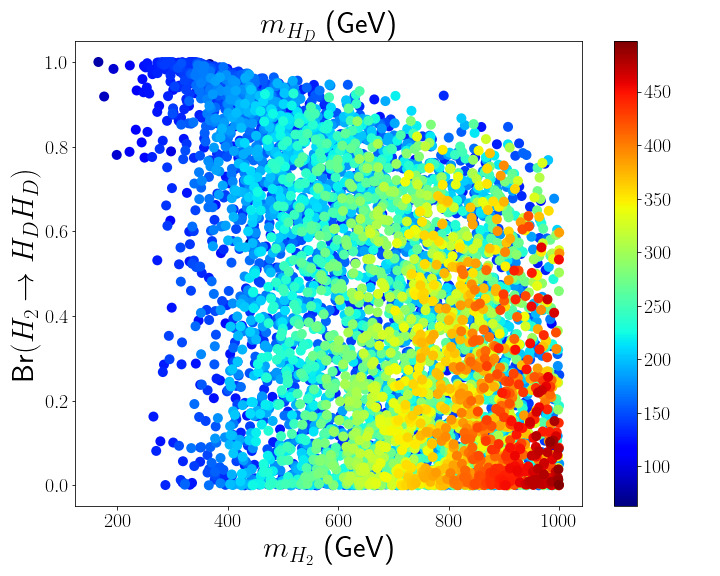}}
    \subfigure[]{\includegraphics[width=0.45\textwidth]{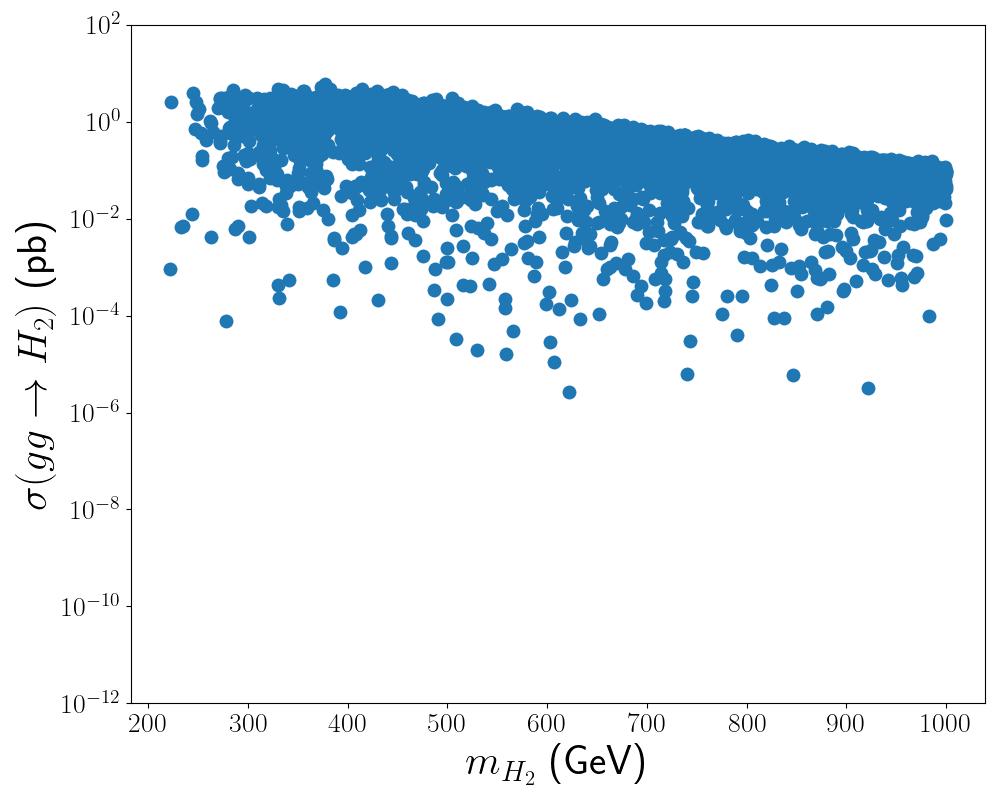}}
    \subfigure[]{\includegraphics[width=0.45\textwidth]{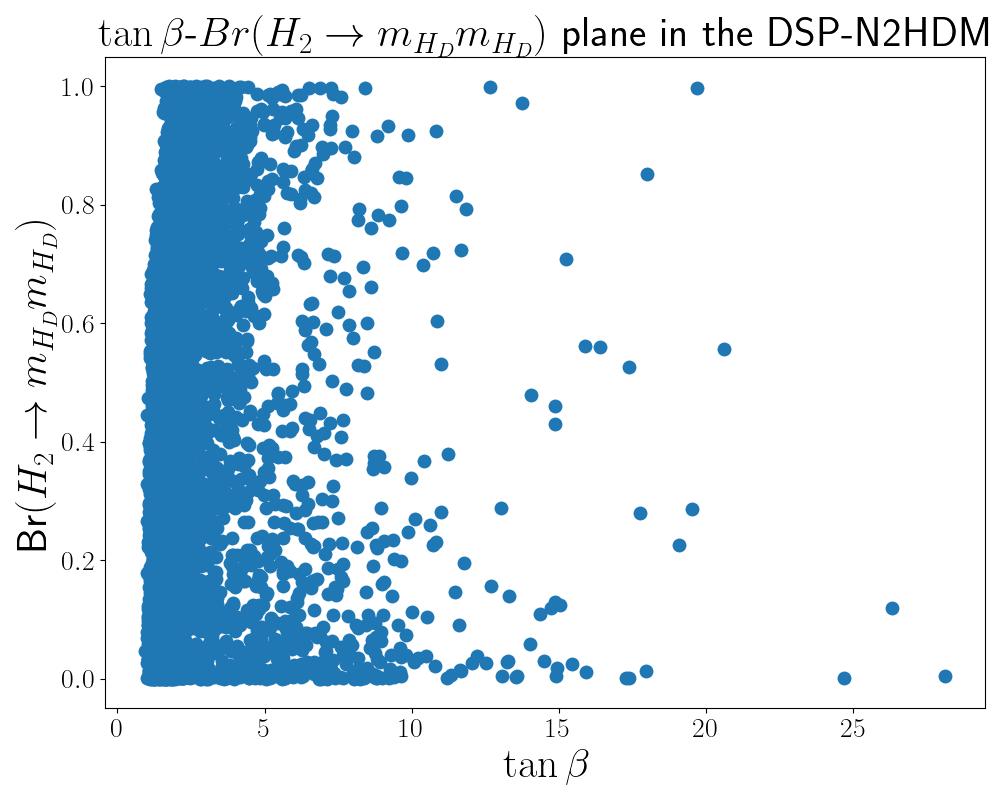}} 
    \subfigure[]{\includegraphics[width=0.45\textwidth]{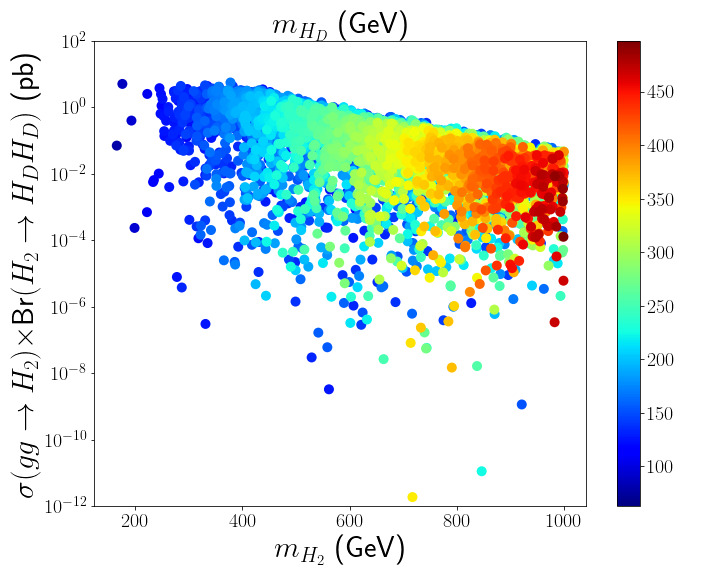}}
    \subfigure[]{\includegraphics[width=0.45\textwidth]{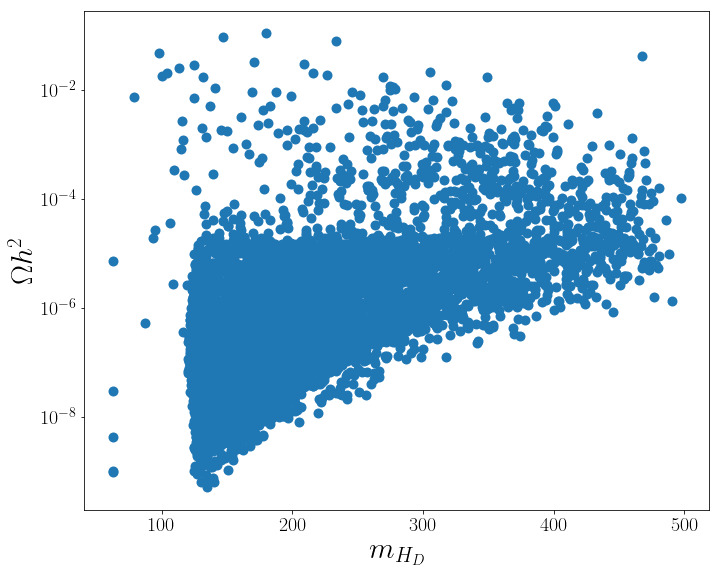}}
    \caption{Two dimensional scatter plots of (a) $Br(H_2 \rightarrow H_D H_D)$ versus $m_{H_2}$ with $m_{H_D}$ shown in color  (b) $\sigma(gg \rightarrow H_2)$ versus $m_{H_2}$ (c) $Br(H_2 \rightarrow H_D H_D)$ versus $\tan{\beta}$ (d) $\sigma \times Br$ versus $m_{H_2}$ with $m_{H_D}$ shown in color (e) The relic density.}
    \label{fig:foobar0}
\end{figure}
The results of the N2HDM with a singlet dark matter are shown in Figures 1.a. to 1.e. We have restricted our analysis to points where $m_{H_2} \leq 1000$ GeV, for which a total of 20k successful points are collected.  
Figure 1.a. shows a scatter plot of the branching ratio of $H_2$ to a pair of dark matter particles ($Br(H_2 \rightarrow H_D H_D)$) versus $m_{H_2}$. The color indicates the dark matter mass, $m_{H_D}$. As can be seen, $m_{H_2}$ ranges from 222 GeV to 1000 GeV, while $m_{H_D}$ ranges from $100$ GeV to $497$ GeV. The branching ratio $Br(H_2 \rightarrow H_D H_D)$ can reach values close to one, especially for regions where the mass of the dark singlet is below 200 GeV. As the mass increases, the branching ratio reduces significantly, as seen in the plot's red corner.
A representative point where $Br(H_2 \rightarrow H_D H_D) \sim 1$ corresponds to the parameter space point: $m_{H_2} = 337$ GeV, $m_{H_D} = 127$ GeV, $\tan{\beta}$ = 3, $\sigma(gg \rightarrow H_2) = 1.1$ pb, and $\Omega h^2 = O(10^{-9})$.

Next, Fig. 1.b. displays the production of $H_2$ via gluon fusion ($\sigma(gg \rightarrow H_2)$) versus $m_{H_2}$. It ranges from $3 \times 10^{-6}$ pb to $5.8$ pb. The maximum value occurs at $m_{H_2} = 378$ GeV, $m_{H_D} = 170$ GeV, $\tan{\beta}$ = 1.5, $Br(H_2 \rightarrow H_D H_D) = 0.95 $, and $\Omega h^2 = O(10^{-8})$. The dependence on $\tan{\beta}$ is shown in Fig. 1.c. First, in the allowed regions, $\tan{\beta}$ takes values between 0.9 and 28. We note that most successful points lie in regions where $\tan{\beta} < 5$. The distribution of the branching ratio to dark matter appears to be nearly uniform. However, smaller values of $\tan{\beta}$ tend to be associated with branching ratio below 0.1. As $\tan{\beta}$ slightly increases from 0.9 to 1.5, the branching ratio exceeds 0.6.  

Fig. 1.d. shows the branching ratio times the cross-section. The values range between $O(10^{-12})$ pb and 5.6 pb. Moreover, for a given mass of $m_{H_2}$, the maximum value of $Br \times \sigma$ slightly decreases with the increase of $m_{H_2}$. The maximum value of $Br \times \sigma$ found in the allowed parameter space is achieved at $m_{H_2} = 378$ GeV, $m_{H_D} = 170$ GeV, $\tan{\beta}$ = 1.5, $Br(H_2 \rightarrow H_D H_D) = 0.95 $, and $\Omega h^2 = O(10^{-8})$. At this point, $H_2$ mainly decays into $t \bar{t}$ with a percentage of $3 \%$, followed by $h_1 h_1$ at 0.5 $\%$ and $W^+ W^-$ at 0.4 $\%$.  

Finally, we note that the relic density in the scanned parameter space is almost always below the lower Planck bound, as shown in Fig. 1.e. Hence, while this parameter space provides a candidate for dark matter, it cannot explain all dark matter phenomena. A representative point where the relic density is fully explained by this model corresponds to $m_{H_2} = 914$ GeV, $m_{H_D} = 180$ GeV, $\tan{\beta}$ = 1.4, $Br(H_2 \rightarrow H_D H_D) \approx 0.11 $, $\sigma(gg \rightarrow H_2) \approx 0.11$ pb and $\Omega h^2 \approx 0.11$. In this case, the dominant decay of $H_2$ is to $t\bar{t}$ pair at 88 $\%$.

\subsection{The NUH-NMSSM}
\label{res:nmssm}
In this subsection, we present the results for the NMSSM with boundary conditions set at the GUT scale. A total of 120k valid points are collected, and the data corresponds to $m_{H_2} < 1$ TeV.  
Figure 2.a shows the results of our scans in the usual $m_0$-$m_{1/2}$ plane, where the first parameter ranges from $\sim 0$ to $5961$ GeV, while the second ranges between $690$ GeV and $10460$ GeV. As can be seen in the Figure, values of $m_{H_2} \leq 350$ GeV correspond to portions of the parameter space where $m_{1/2} < 2500$ GeV, while $m_0$ can take the full range of its allowed values. Regions where $m_{1/2} > 2500$ GeV are associated with $m_{H_2} > 350$ GeV. The shape of the parameter space reflects the fact that we have used both random scanning method (especially the box where $m_{1/2}, m_0 < 2500$ GeV) as well as the MCMC method described in Sec.~\ref{sec:scans}.

As mentioned in Sec.~\ref{intro}, the main production channel of the neutral scalars is via the fusion of two gluons. The results are displayed in Figure 2.b where the minimum and the maximum values of the production as a function of $m_{H_2}$ are $\mathcal{O}(10^{-10})$ pb and 2.2 pb, respectively. In the subsequent analysis, we require that $Br(H_2 \rightarrow \tilde{\chi}_1 \tilde{\chi}_1) > 0$. By doing so, some points in the parameter space are removed, and what remains is a parameter space where $\sigma^{\text{max}}_{gg \rightarrow H_2} \sim 0.51$ pb. This corresponds to $m_{H_2} = 210$ GeV, $m_{\tilde{\chi}_1} = 101$ GeV, and $Br(H_2 \rightarrow \tilde{\chi}_1 \tilde{\chi}_1)) = 0.012$. 

Moving on to the branching ratio, Figure 2.c depicts the allowed ranges of $Br(H_2 \rightarrow \tilde{\chi}_1 \tilde{\chi}_1) $ as a function of $m_{H_2}$ and $m_{\tilde{\chi}_1}$ (indicated by color).
The branching ratio takes values ranging from $\sim 0$ to $0.87$. The highest value occurs at $m_{H_2} = 212$ GeV, and $m_{\tilde{\chi}_1} = 98$ GeV. These values correspond to $\tan{\beta} = 17$, and are associated with $\Omega h^2 \sim 0.01$ and $\sigma_{ggH_2} \sim 0.1$ pb. In this case, $H_2$ decays to $b \bar{b}$ with  Br$\sim 0.1$, and to $W^+W^-$ and $\tau \bar{\tau}$ with Br$\sim 0.01$.

Figure 2.d shows the dependence of the invisible decay branching ratio on $\tan{\beta}$, ranging from 1.8 to 30. The maximum value of $Br(H_2 \rightarrow \tilde{\chi}_1 \tilde{\chi}_1) $ is obtained only for $\tan{\beta} > 5$.

Figure 2.e displays $\sigma(gg \rightarrow H_2) \times Br(H_2 \rightarrow \tilde{\chi}_1 \tilde{\chi}_1)$. The maximum value is 0.1 pb, which takes place at $m_{H_2} = 216$ GeV, $m_{\tilde{\chi}_1} = 102$ GeV, $\tan{\beta} = 12$, $\Omega h^2 = 0.001$, $Br \approx 0.3$ and $\sigma \approx 0.3$ pb. In this case, $H_2$ is more likely to decay into $W^+W^-$, which accounts for $50\%$ of the decays, while $ZZ$ accounts for $20\%$.

For completeness, Figure 2.f shows the relic density as a function of the mass of the DM particle. All points in the parameter space are associated with values below the lower Planck limit, in which case the DM particle is insufficient to account for all of the observed DM relic density. 

We turn to the case where $\tilde{\chi}_2$ and $\tilde{\chi}_1$ are mass degenerate. This interesting case can lead to $\tilde{\chi}_2$ being long-lived and hence escaping detection. There are two cases here. The first is
when $H_2$ decays to $\tilde{\chi}_2$ and $\tilde{\chi}_1$, and the second is when it decays to a pair of $\tilde{\chi}_2$. The results for both cases are shown in Figure 3. Starting with $H_2 \rightarrow \tilde{\chi}_2 \tilde{\chi}_1$, we can see in Figure 3.a that the branching ratio can reach a maximum value of 0.01. Here the most likely decays are to $\tilde{\chi}_1 \tilde{\chi}_1$ at $45\%$, to $\chi^+ \chi^-$ at $40\%$, to to $W^+ W^-$ at $6.5\%$, to $b \bar{b}$ at $3\%$, to $ZZ$ at $3\%$ , and to $\tilde{\chi}_2 \tilde{\chi}_2$ at $2\%$.  
Furthermore, the maximum value of $\sigma \times Br$ is 0.001 pb, which is displayed in Figure 3.c. This occurs at $m_{H_2} = 456$ GeV, $m_{\tilde{\chi}_1} \approx m_{\tilde{\chi}_2} = 107$ GeV, $\sigma(gg \rightarrow H_2) \sim 0.23$ pb, and $Br(H_2 \rightarrow \tilde{\chi}_2 \tilde{\chi}_1) \sim 0.003$. The other relevant decays of $H_2$ account for branching ratios of $0.48$ to $b\bar{b}$, $0.35$ to $t\bar{t}$, $0.065$ to $\tau \bar{\tau}$, $0.041$ to $H_1 H_1$, and $0.02$ to $W^+ W^-$ and $\chi^+ \chi^-$.

Finally, we consider the second case, displayed in Figure 3.b. The maximum value of $Br(H_2 \rightarrow \tilde{\chi}_2 \tilde{\chi}_2)$ is $0.25$. while it is $0.5$ to $\chi^+ \chi^-$, and $0.25$ to $\tilde{\chi}_1 \tilde{\chi}_1$. And the maximum value of $\sigma \times Br$ is 0.01 pb, which is achieved at $m_{H_2} = 291$ GeV, $m_{\tilde{\chi}_1} \approx m_{\tilde{\chi}_2} = 109$ GeV, $\sigma(gg \rightarrow H_2) \sim 0.13$ pb, and $Br(H_2 \rightarrow \tilde{\chi}_2 \tilde{\chi}_1) \sim 0.1$. Other relevant branching ratios of $H_2$ are $0.29$ to $W^+ W^-$, $0.2$ to $\chi^+ \chi^-$, $0.18$ to $H_1 H_1$, $0.14$ to both $\tilde{\chi}_1 \tilde{\chi}_1$, and $0.12$ to $ZZ$.

\begin{figure}[!ht]
    \centering
    \subfigure[]{\includegraphics[width=0.45\textwidth]{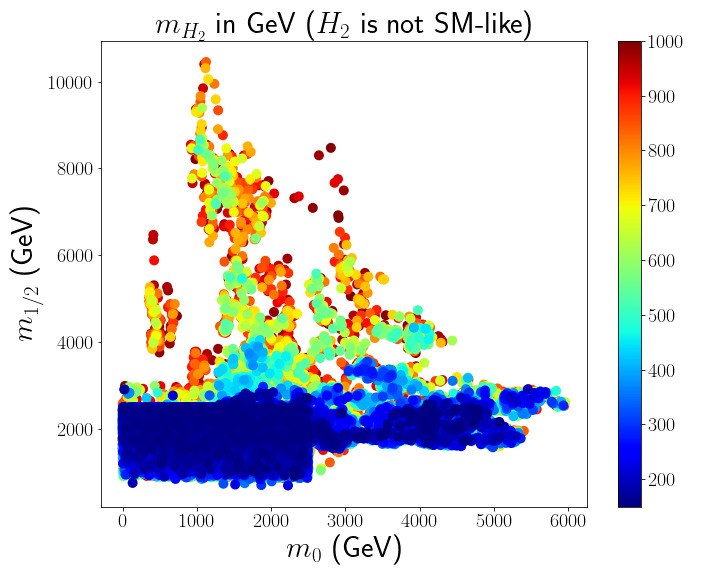}}
    \subfigure[]{\includegraphics[width=0.45\textwidth]{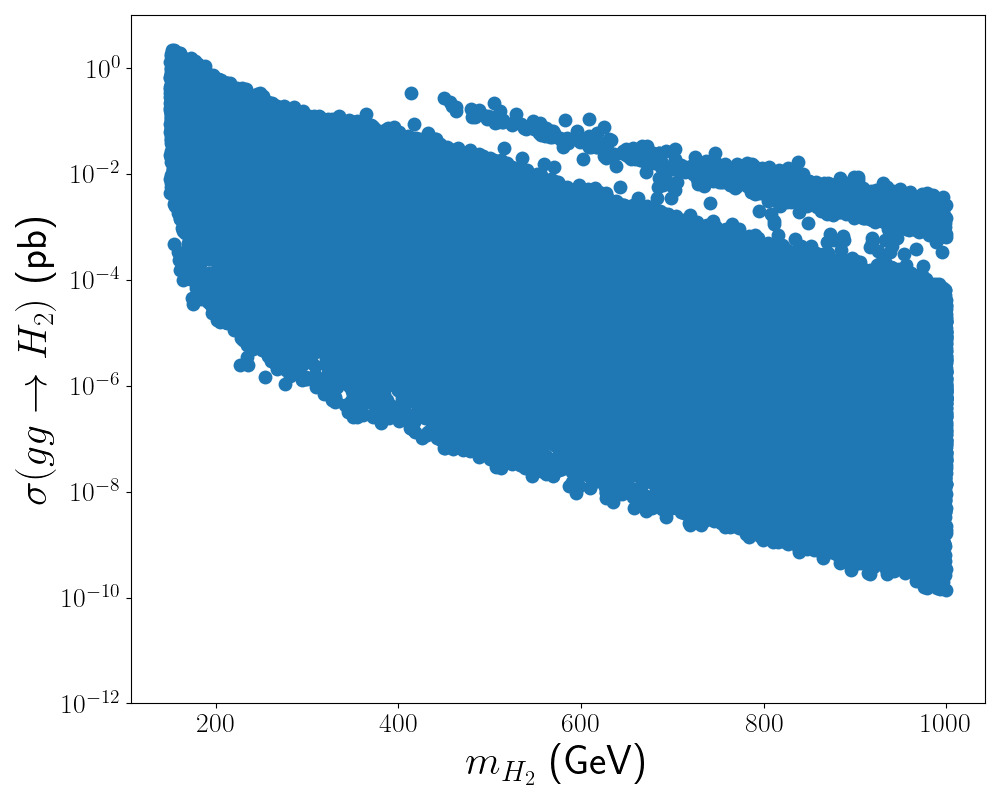}}
    \subfigure[]{\includegraphics[width=0.45\textwidth]{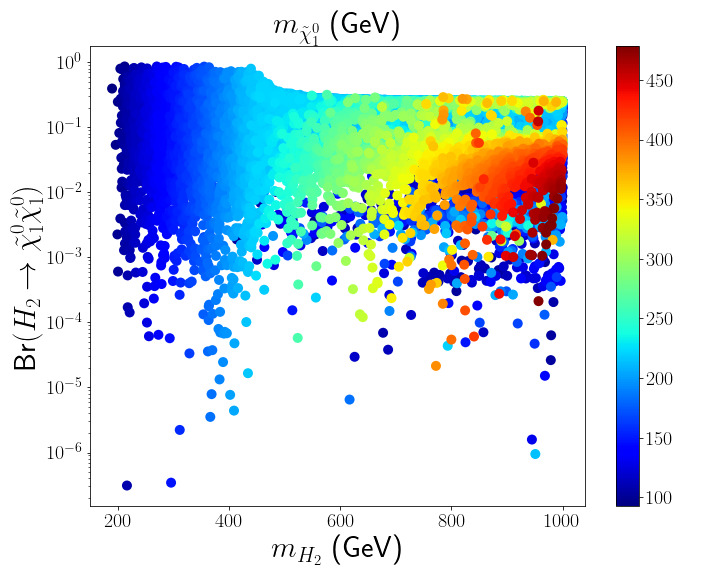}}
    \subfigure[]{\includegraphics[width=0.45\textwidth]{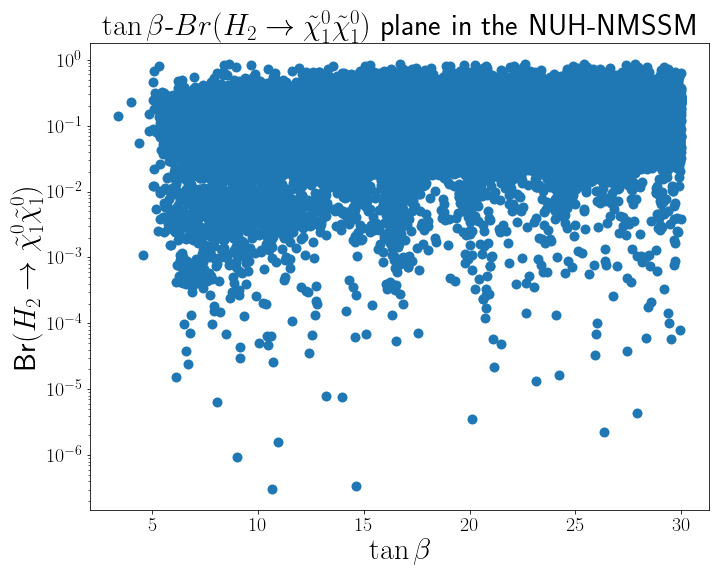}}
    \subfigure[]{\includegraphics[width=0.45\textwidth]{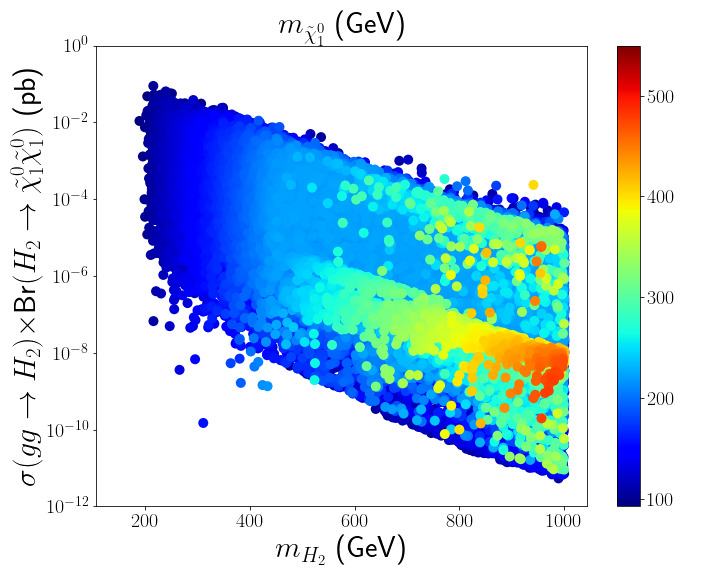}}  
    \subfigure[]{\includegraphics[width=0.45\textwidth]{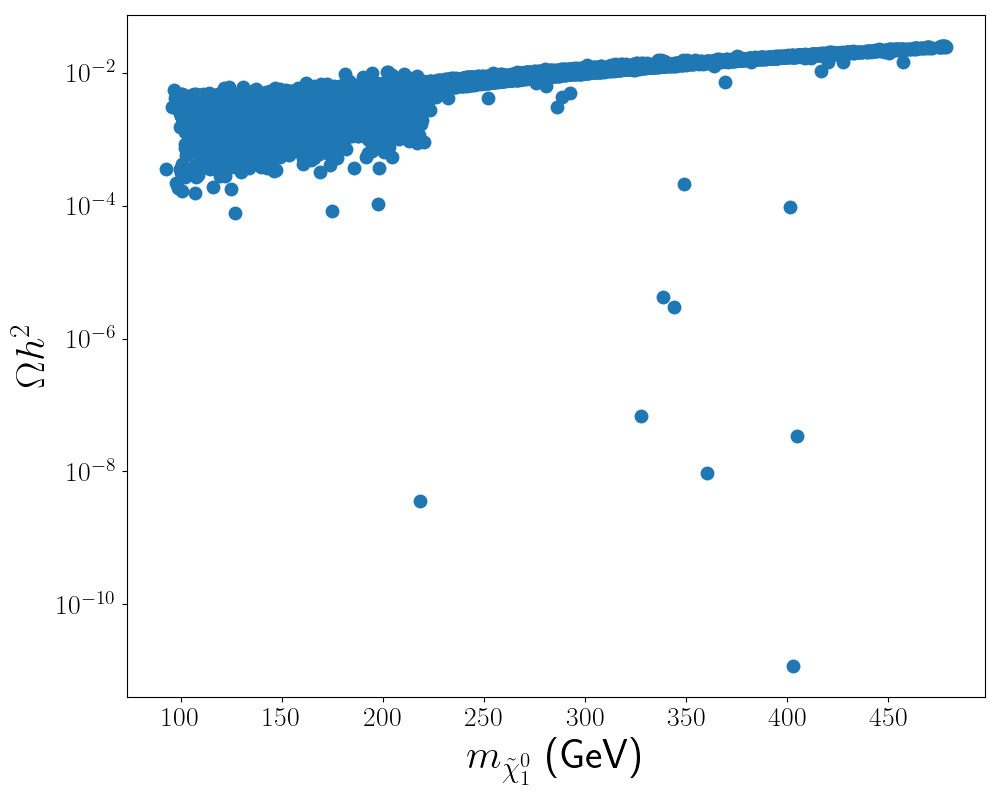}}  
    \caption{Two dimensional scatter plots of (a) $m_{1/2}$ versus $m_0$ (b) $\sigma(gg \rightarrow H_2)$ as a function of $m_{H_2}$ (c) $Br(H_2 \rightarrow \tilde{\chi}_1 \tilde{\chi}_1 )$ as a function of $m_{H_2}$ (d) The dependence of $Br$ on $\tan{\beta}$ (e) $\sigma \times Br$ as a function of $m_{H_2}$ (f) The relic density.}
    \label{fig:foobar1}
\end{figure}

\begin{figure}[!ht]
    \centering
    \subfigure[]{\includegraphics[width=0.45\textwidth]{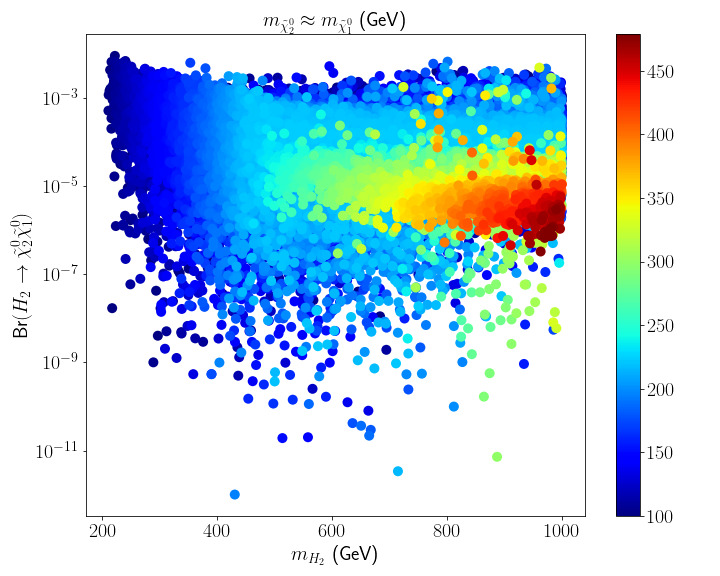}} 
    \subfigure[]{\includegraphics[width=0.45\textwidth]{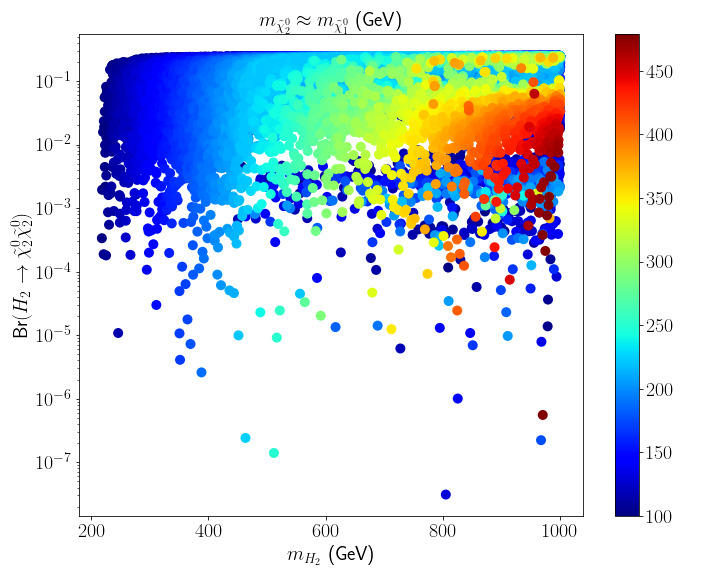}} 
    \subfigure[]{\includegraphics[width=0.45\textwidth]{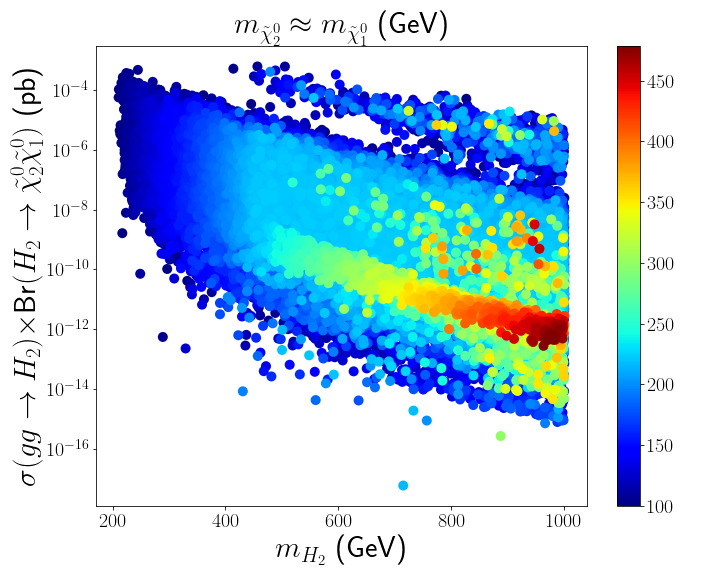}}
    \subfigure[]{\includegraphics[width=0.45\textwidth]{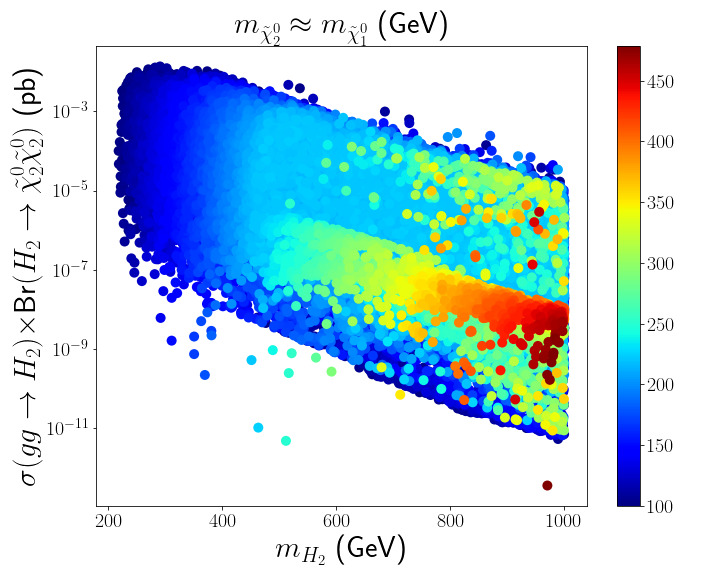}}    
    \caption{Two dimensional scatter plots of (a) $Br(H_2 \rightarrow \tilde{\chi}_2 \tilde{\chi}_1)$ versus $m_{H_2}$ (b) $Br(H_2 \rightarrow \tilde{\chi}_2 \tilde{\chi}_2)$ versus $m_{H_2}$  (c) $\sigma \times Br(H_2 \rightarrow \tilde{\chi}_2 \tilde{\chi}_1)$ versus $m_{H_2}$  (d) $\sigma \times Br(H_2 \rightarrow \tilde{\chi}_2 \tilde{\chi}_2)$ versus $m_{H_2}$ }
    \label{fig:foobar2}
\end{figure}

\section{Discussion and Conclusion} \label{discon}
In this paper, we considered the production of a non-SM-like CP-even Higgs boson $H_2$ via gluon fusion at 13 TeV, and its subsequent decay into DM particle in two well-motivated extensions of the SM, the DSP-N2HDM, and the NUH-NMSSM. 

The parameter space of the DSP-N2HDM was scanned subject to recent experimental constraints, including LHC and DM searches. The analysis was restricted to points where $m_{H_2} \leq 1000$ GeV. We found that the branching ratio $Br(H_2 \rightarrow H_D H_D)$ can reach values close to one, especially for regions where the mass of the dark singlet is below 200 GeV, which is consistent with previous literature. The production of $H_2$ via gluon fusion ranges between $3 \times 10^{-6}$ pb and $5.8$ pb, with most successful points having $\tan{\beta} < 5$. The maximum value of $Br \times \sigma$ found in the allowed parameter space corresponds to $m_{H_2} = 378$ GeV, $m_{H_D} = 170$ GeV, $\tan{\beta}$ = 1.5, $Br(H_2 \rightarrow H_D H_D) = 0.95 $, and $\Omega h^2 = O(10^{-8})$. We note that the relic density is almost always below the lower bound. Hence, this region of parameter space is insufficient to explain all dark matter phenomena.

On the other hand, the allowed parameter space of the NUH-NMSSM (with $m_{H_2} < 1$ TeV) was also explored under recent experimental limits. The production cross-section at $\sqrt{s} = 13$ TeV was computed, with values ranging from $\mathcal{O}(10^{-10})$ pb to 2.2 pb, and only the points with $Br(H_2 \rightarrow \tilde{\chi}_1 \tilde{\chi}_1) > 0$ were considered. The allowed range of the branching ratio $Br(H_2 \rightarrow \tilde{\chi}_1 \tilde{\chi}_1)$ was computed, and it assumes values between $\sim 0$ to $0.87$. The maximum value of the quantity $\sigma(gg \rightarrow H_2) \times Br(H_2 \rightarrow \tilde{\chi}_1 \tilde{\chi}_1)$ was found to be 0.1 pb. 

In the case where $\tilde{\chi}_2$ and $\tilde{\chi}_1$ are mass degenerate, there are two cases to consider. Firstly, when $H_2$ decays to $\tilde{\chi}_2$ and $\tilde{\chi}_1$, the branching ratio can reach a maximum value of 0.01, with the most likely decays being to $\tilde{\chi}_1 \tilde{\chi}_1$ and $\chi^+ \chi^-$. The maximum value of $\sigma \times Br$ is 0.001 pb, occurring when $m_{H_2} = 456$ GeV, $m_{\tilde{\chi}_1} \approx m_{\tilde{\chi}_2} = 107$ GeV, $\sigma(gg \rightarrow H_2) \sim 0.23$ pb, and $Br(H_2 \rightarrow \tilde{\chi}_2 \tilde{\chi}_1) \sim 0.003$. The second case is when $H_2$ decays to a pair of $\tilde{\chi}_2$, with a maximum branching ratio of $0.25$ to $\tilde{\chi}_2 \tilde{\chi}_2$, and $0.5$ to $\chi^+ \chi^-$ and $\tilde{\chi}_1 \tilde{\chi}_1$. The maximum value of $\sigma \times Br$ is 0.01 pb, occurring at $m_{H_2} = 291$ GeV, $m_{\tilde{\chi}_1} \approx m_{\tilde{\chi}_2} = 109$ GeV, $\sigma(gg \rightarrow H_2) \sim 0.13$ pb, and $Br(H_2 \rightarrow \tilde{\chi}_2 \tilde{\chi}_1) \sim 0.1$. The relevant decay branching ratios of $H_2$ were also presented.

The results obtained from the analysis of the DPS-N2HDM and NUH-NMSSM parameter spaces showed the accommodated values of the production of a non-SM-like $H_2$ via gluon fusion and its subsequent decay into DM particles. However, the DPS-N2HDM allowed for higher production cross-sections than the NUH-NMSSM, with the maximum value of $Br \times \sigma$ found in the DPS-N2HDM being larger than that in the NUH-NMSSM. Moreover, the allowed parameter space in the DPS-N2HDM showed that the branching ratio $Br(H_2 \rightarrow H_D H_D)$ reaches values close to $100\%$, especially for regions where the mass of the dark singlet is below 200 GeV. On the other hand, the NUH-NMSSM parameter space allowed for quasi-invisible decays where the LSP and NLSP are mass-degenerate. In both models, the relic density was found to be below the lower bound, indicating that this parameter space can partially explain dark matter. And while the DM particle is an SM-singlet scalar in the DSP-N2HDM, it was mostly Higgsino in the scanned parameter space of the NUH-NMSSM. It is worth emphasizing that in the MSSM, the parameter space with Higgsino-like DM is largely restricted compared with the NMSSM. Indeed, it was shown in Ref.~\cite{Ellis:2022emx} that the viable regions of the constrained MSSM parameter space allow for Higgsino DM with a mass ranging from 1 TeV to 1.1 TeV.

In conclusion, we showed that the DSP-N2HDM and NUH-NMSSM can pass all recent experimental constraints
while providing challenging scenarios for collider searches of extended Higgs sectors, focusing particularly on a non-SM-like $H_2$. We also showed that the predictions of these models for the invisible decay of $H_2$ to dark matter particles are quite different, indicating a possible way to distinguish both models. We made no assumptions regarding the relationship between the N2HDM and the NMSSM. 

\section*{Acknowledgement}
MB acknowledges support from the King Saud University Deanship of Scientific Research.


\end{document}